\documentstyle[12pt]{article}
\newlength{\extraspace}
\newlength{\extraspaces}
\newcommand{\be}{\begin{equation}
\addtolength{\abovedisplayskip}{\extraspaces}
\addtolength{\belowdisplayskip}{\extraspaces}
\addtolength{\abovedisplayshortskip}{\extraspace}
\addtolength{\belowdisplayshortskip}{\extraspace}}
\newcommand{\ee}{\end{equation}}
\newcommand{\ba}{\begin{eqnarray}
\addtolength{\abovedisplayskip}{\extraspaces}
\addtolength{\belowdisplayskip}{\extraspaces}
\addtolength{\abovedisplayshortskip}{\extraspace}
\addtolength{\belowdisplayshortskip}{\extraspace}}
\newcommand{\ea}{\end{eqnarray}}
\newcommand{\nonu}{\nonumber \\[.5mm]}
\newcommand{\A}{&\!\!\!}
\begin{document}
\thispagestyle{empty}
\begin{flushright}
SIT-LP-04/06 \\
{\tt hep-th/0406182} \\
June, 2004
\end{flushright}
\vspace{7mm}
\begin{center}
{\large \bf More on the universality of the Volkov-Akulov action 
under $N = 1$ nonlinear supersymmetry 
} \\[20mm]
{\sc Kazunari Shima}
\footnote{
\tt e-mail: shima@sit.ac.jp} \ 
and \ 
{\sc Motomu Tsuda}
\footnote{
\tt e-mail: tsuda@sit.ac.jp} 
\\[5mm]
{\it Laboratory of Physics, 
Saitama Institute of Technology \\
Okabe-machi, Saitama 369-0293, Japan} \\[20mm]
\begin{abstract}
We discuss further the universality of the Volkov-Akulov (V-A) action 
of a Nambu-Goldstone (N-G) fermion 
for the spontaneous breaking of supersymmetry (SUSY). 
We show general relations between the standard V-A action 
and nonlinear (NL) SUSY actions including apparently 
(pathological) higher derivatives of the N-G fermion. 
Composite fields of the N-G fermions are found, 
which transform homogeneously under NL SUSY transformations of V-A. 
Consequently, we obtain NL SUSY invariant constraints 
which connect our NL SUSY actions with the V-A action. 
The constraints are explicitly solved and we show examples 
of the NL SUSY actions which are equivalent to the V-A action. 

/Keywords: supersymmetry, Nambu-Goldstone fermion 
\end{abstract}
\end{center}

\newpage

Nonlinear (NL) realization of supersymmetry (SUSY) 
and NL SUSY action given by Volkov-Akulov (V-A) \cite{VA} 
are decribed in terms of a Nambu-Goldstone (N-G) 
fermion \cite{SS} indicating the spontaneous SUSY breaking \cite{FI,O}. 
The equivalence of the V-A model of NL SUSY 
to various linear (L) supermultiplets \cite{WZ,Fa} was shown 
by many authors \cite{IK}-\cite{STT}. 
In the relation between the NL and the L SUSY, 
basic fields in the L supermultiplets are expressed 
as composites of the N-G fermion in SUSY invariant way, 
and this fact gives deep insight 
towards the unification of spacetime and matter 
from the viewpoint of compositeness of matter 
as discussed in \cite{KSST}. 
While, it is known that there exists a nontrivial 
NL SUSY higher derivative action of the N-G fermion 
as is exemplified in \cite{SW}. 
In order to understand the implications of the nontrivial 
higher derivative action of the N-G fermion, 
it is useful to investigate the relation among NL SUSY actions 
i.e., the universality of NL SUSY actions with the N-G fermion. 
Recently, this problem has been discussed in \cite{HK} 
in the viewpoint of the braneworld scenario. 
And by heuristic arguments we have discussed in \cite{ST} 
the relation between the standard V-A action 
and a NL SUSY action including apparently a (Weyl) ghost field 
which originates from pathological higher derivatives of the N-G fermion. 

In this letter, by extending the arguments of \cite{ST} 
with respect to the universality of NL SUSY actions 
to the cases where the order of derivatives of the N-G fermion 
is higher than in \cite{ST}, 
we discuss more general relations 
between the standard V-A action and NL SUSY actions 
including apparently (pathological) higher derivatives 
of the N-G fermion under $N = 1$ NL SUSY. 
By using the algorithmic procedure given in \cite{Iv}, 
we find composite fields of the N-G fermions 
which transform homogeneously under NL SUSY transformations of V-A. 
Consequently, we obtain NL SUSY invariant constraints 
which connect our NL SUSY actions with the standard V-A action. 
The constraints are explicitly solved and we show examples 
of the NL SUSY actions which are equivalent to the V-A action. 
We show that the arguments of \cite{ST} 
with respect to the universality of NL SUSY actions 
are derived as a solution of the constraints in the examples. 

Let us begin with the brief review of the NL realization of SUSY 
by V-A \cite{VA}. In the $N = 1$ V-A model, 
the NL SUSY transformation law of a (Majorana) N-G fermion 
$\psi$ generated by a constant (Majorana) spinor parameter $\zeta$ is 
\footnote{
In this letter Minkowski spacetime indices are denoted 
by $a, b, ... = 0, 1, 2, 3$, 
and we use the  Minkowski spacetime metric 
${1 \over 2}\{ \gamma^a, \gamma^b \} = \eta^{ab}= (+, -, -, -)$ 
and $\sigma^{ab} = {i \over 4}[\gamma^a, \gamma^b]$.}, 
\be
\delta_Q \psi = {1 \over \kappa} \zeta 
- i \kappa (\bar\zeta \gamma^a \psi) \partial_a \psi, 
\label{NLSUSY}
\ee
where $\kappa$ is a constant whose dimension is $({\rm mass})^{-2}$. 
The NL SUSY transformation (\ref{NLSUSY}) satisfies 
a closed off-shell commutator algebra, 
$[\delta_Q(\zeta_1), \delta_Q(\zeta_2)]$ $= \delta_P(v)$, 
where $\delta_P$ is a translation with a parameter 
$v^a = 2i \bar\zeta_1 \gamma^a \zeta_2$. 
Based on the invariant one-form under Eq.(\ref{NLSUSY}), 
i.e., $\omega^a = w^a{}_b dx^a 
= (\delta^a_b - i \kappa^2 \bar\psi \gamma^a \partial_b \psi) dx^a$, 
the NL SUSY V-A action $S_{\rm VA}(\psi)$ is given by 
\ba
S_{\rm VA}(\psi) = \A \A - {1 \over {2 \kappa^2}} 
\int d^4 x \ \vert w \vert \nonu
= \A \A 
- {1 \over {2 \kappa^2}} \int d^4 x 
\left[ 1 + t{^a}_a 
+ {1 \over 2}(t{^a}_a t{^b}_b - t{^a}_b t{^b}_a) \right. \nonu
\A \A 
\left. - {1 \over 6} \epsilon_{abcd} \epsilon^{efgd} t{^a}_e t{^b}_f t{^c}_g 
- {1 \over 4!} \epsilon_{abcd} \epsilon^{efgh} t{^a}_e t{^b}_f t{^c}_g t{^d}_h 
\right], 
\label{VAact}
\ea
where $\vert w \vert = \det w^a{}_b$ 
and $t^a{}_b = - i \kappa^2 \bar\psi \gamma^a \partial_b \psi$. 

On the other hand, let us consider NL SUSY actions 
which include (pathological) higher derivatives 
of a N-G fermion in addition to the standard V-A action 
as exemplified in \cite{ST}. 
Namely, we denote $\lambda$ for the (Majorana) N-G fermion 
which transforms into $\psi$ in Eq.(\ref{VAact}) 
through NL SUSY invariant constraints as will be shown later, 
and we propose the actions $S(\lambda)$ including 
apparently nontrivial terms with (pathological) higher derivatives 
of $\lambda$, 
\be
S(\lambda) = S_{\rm VA}(\lambda) + [\ {\rm higher\ derivative\ terms\ of\ } \lambda\ ]
\label{NLSUSYact}
\ee
which are invariant under the NL SUSY transformation of $\lambda$, 
\be
\delta_Q \lambda = {1 \over \kappa} \zeta 
- i \kappa (\bar\zeta \gamma^a \lambda) \partial_a \lambda. 
\label{NLSUSY2}
\ee
Note that the form of Eq.(\ref{NLSUSY2}) is the same as Eq.(\ref{NLSUSY}). 

In order to construct the NL SUSY invariant constraints 
between the N-G fermions $\psi$ and $\lambda$,  
we use the algorithmic procedure given by Ivanov \cite{Iv} 
to pass to a relevant NL SUSY theory from another one. 
Indeed, first we introduce the fields, 
\be
\lambda + \sum_{n \ge 1} c_n 
\ (i \kappa^{1 \over 2})^n \gamma^A \partial_A \lambda, 
\label{fields}
\ee
as the most general form of ${\cal O}(\lambda^1)$ 
in terms of $\lambda$ and its higher derivatives 
with the arbitrary coefficients $c_n$, 
where $\gamma^A$ and $\partial_A$ are defined 
respectively as 
\ba
\A \A \gamma^A = \prod_{\alpha=1}^n \gamma^{a_\alpha} 
= \gamma^{a_1} \gamma^{a_2} \cdots \gamma^{a_n}, 
\nonu
\A \A \partial_A = \prod_{\alpha=1}^n \partial_{a_\alpha} 
= \partial_{a_1} \partial_{a_2} \cdots \partial_{a_n} 
\ea
with $a_1, a_2, ..., a_n$ being Minkowski spacetime indices, 
and $\gamma^A \partial_A$ means $\gamma^A \partial_A = \ \!\not\!\partial^n$. 
Second we show explicitly the following finite transformations 
of the fields (\ref{fields}) for the simplified case of $c_n = 1$, 
\ba
\tilde \lambda(\zeta) 
\A = \A \left( 1 + \delta_\zeta + {1 \over 2!} \delta_\zeta^2 
+ {1 \over 3!} \delta_\zeta^3 + {1 \over 4!} \delta_\zeta^4 \right) 
\left\{ \ \lambda 
+ \sum_{n \ge 1} 
\ (i \kappa^{1 \over 2})^n \gamma^A \partial_A \lambda \ \right\} 
\nonu
\A = \A e^{\delta_\zeta} 
\ \left\{ \ \lambda 
+ \sum_{n \ge 1} 
\ (i \kappa^{1 \over 2})^n \gamma^A \partial_A \lambda \ \right\} 
\label{tilde-z}
\ea
which are generated by the NL SUSY transformations (\ref{NLSUSY2}). 
Note that in Eq.(\ref{tilde-z}) the terms higher than $\delta_\zeta^4$ 
vanish by means of $\zeta^n = 0$ for $n \ge 5$. 
By replacing the spinor parameter $\zeta$ in Eq.(\ref{tilde-z}) 
by $- \kappa \psi$, we finally define the composite fields, 
\be
\tilde \lambda(\psi) = \tilde \lambda_0(\psi) + \tilde \lambda_1(\psi), 
\ee
where $\tilde \lambda_0(\psi)$ and $\tilde \lambda_1(\psi)$ 
are the fields for the finite transformation of $\lambda$ 
and its higher derivatives, respectively, 
i.e., 
\ba
\A \A \tilde \lambda_0(\psi) 
= \left( 1 + \delta_\zeta + {1 \over 2!} \delta_\zeta^2 
+ {1 \over 3!} \delta_\zeta^3 + {1 \over 4!} \delta_\zeta^4 \right) 
\ \lambda \ \ \vert_{\zeta \rightarrow - \kappa \psi}, 
\label{tilde-0}
\\
\A \A \tilde \lambda_1(\psi) 
= \left( 1 + \delta_\zeta + {1 \over 2!} \delta_\zeta^2 
+ {1 \over 3!} \delta_\zeta^3 + {1 \over 4!} \delta_\zeta^4 \right) 
\ \sum_{n \ge 1} 
\ (i \kappa^{1 \over 2})^n \gamma^A \partial_A \lambda 
\ \ \vert_{\zeta \rightarrow - \kappa \psi}. 
\label{tilde-1}
\ea
The explicit form of $\tilde \lambda_0(\psi)$ becomes 
\ba
\A \A \tilde \lambda_0(\psi) 
\nonu
\A \A = \lambda - \psi + \eta^a(\psi) \ \partial_a \lambda 
+ i \kappa^2 \eta^a(\psi) \ \bar\psi \gamma^b \partial_a \lambda 
\ \partial_b \lambda 
+ {1 \over 2} \eta^a(\psi) \eta^b(\psi) \ \partial_a \partial_b \lambda 
\nonu
\A \A 
- \kappa^4 \eta^a(\psi) \ \bar\psi \gamma^b \partial_a \lambda 
\ \bar\psi \gamma^c \partial_b \lambda \ \partial_c \lambda 
+ {i \over 2} \kappa^2 \eta^a(\psi) \eta^b(\psi) 
\ \bar\psi \gamma^c \partial_a \partial_b \lambda \ \partial_c \lambda 
\nonu
\A \A 
+ i \kappa^2 \eta^a(\psi) \eta^b(\psi) 
\ \bar\psi \gamma^c \partial_a \lambda \ \partial_b \partial_c \lambda 
+ {1 \over 6} \eta^a(\psi) \eta^b(\psi) \eta^c(\psi) 
\ \partial_a \partial_b \partial_c \lambda 
\nonu
\A \A 
- i \kappa^6 \eta^a(\psi) \ \bar\psi \gamma^b \partial_a \lambda 
\ \bar\psi \gamma^c \partial_b \lambda 
\ \bar\psi \gamma^d \partial_c \lambda \ \partial_d \lambda 
- \kappa^4 \eta^a(\psi) \eta^b(\psi) 
\ \bar\psi \gamma^c \partial_a \lambda 
\ \bar\psi \gamma^d \partial_b \partial_c \lambda \ \partial_d \lambda 
\nonu
\A \A 
- \kappa^4 \eta^a(\psi) \eta^b(\psi) 
\ \bar\psi \gamma^c \partial_a \lambda 
\ \bar\psi \gamma^d \partial_c \lambda \ \partial_b \partial_d \lambda 
- {1 \over 2} \kappa^4 \eta^a(\psi) \eta^b(\psi) 
\ \bar\psi \gamma^c \partial_a \partial_b \lambda 
\ \bar\psi \gamma^d \partial_c \lambda \ \partial_d \lambda 
\nonu
\A \A 
- {1 \over 2} \kappa^4 \eta^a(\psi) \eta^b(\psi) 
\ \bar\psi \gamma^c \partial_a \lambda 
\ \bar\psi \gamma^d \partial_b \lambda \ \partial_c \partial_d \lambda 
+ {i \over 2} \kappa^2 \eta^a(\psi) \eta^b(\psi) \eta^c(\psi) 
\ \bar\psi \gamma^d \partial_a \partial_b \lambda 
\ \partial_c \partial_d \lambda 
\nonu
\A \A 
+ {i \over 2} \kappa^2 \eta^a(\psi) \eta^b(\psi) \eta^c(\psi) 
\ \bar\psi \gamma^d \partial_a \lambda 
\ \partial_b \partial_c \partial_d \lambda 
+ {i \over 6} \kappa^2 \eta^a(\psi) \eta^b(\psi) \eta^c(\psi) 
\ \bar\psi \gamma^d \partial_a \partial_b \partial_c \lambda 
\ \partial_d \lambda 
\nonu
\A \A 
+ {1 \over 24} \eta^a(\psi) \eta^b(\psi) 
\eta^c(\psi) \eta^d(\psi) 
\ \partial_a \partial_b \partial_c \partial_d \lambda 
\label{tilde-e0}
\ea
with $\eta^a(\psi) = i \kappa^2 \bar\psi \gamma^a \lambda$. 
On the other hand, in order to derive the explicit form 
of $\tilde \lambda_1(\psi)$, we use the Leipniz rule of derivatives 
for some products $A_1 A_2 \cdots A_m$, 
\ba
\partial_A (A_1 A_2 \cdots A_m) 
= \A \A \prod_{\alpha=1}^n \partial_{a_\alpha} (A_1 A_2 \cdots A_m) 
\nonu
= \A \A 
\sum_{k_1+k_2+\cdots+k_m=n} {n! \over {k_1!k_2! \cdots k_m!}} 
\nonu
\A \A 
\times \prod_{\alpha=1}^{k_1} \partial_{a_\alpha} A_1 
\prod_{\beta=k_1+1}^{k_1+k_2} \partial_{a_\beta} A_2 \cdots 
\prod_{\gamma=k_1+k_2+\cdots+k_{m-1}+1}^n \partial_{a_\gamma} A_m, 
\label{Leipniz}
\ea
where $k_j \ge 0$ ($j = 1, 2, ... , m$) 
and the indices $a_\alpha, a_\beta, ... , a_\gamma$ 
are totally symmetrized. 
According to Eq.(\ref{Leipniz}), $\tilde \lambda_1(\psi)$ is given by 
\ba
\A \A \tilde \lambda_1(\psi) 
\nonu
\A \A 
= \sum_{n \ge 1} i^n \kappa^{n \over 2} \gamma^A \partial_A \lambda 
\nonu
\A \A 
+ \sum_{n \ge 1} i^{n+1} \kappa^{{n \over 2} + 2} \gamma^A 
\sum_{k_1+k_2=n} {n! \over {k_1!k_2!}} 
\ \prod_\alpha \bar\psi \gamma^b \partial_{a_\alpha} \lambda 
\prod_\beta \partial_{a_\beta} \partial_b \lambda 
\nonu
\A \A 
- \sum_{n \ge 1} i^n \kappa^{{n \over 2} + 4} \gamma^A 
\sum_{k_1+k_2+k_3=n} {n! \over {k_1!k_2!k_3!}} 
\nonu
\A \A 
\times \left( 
\prod_\alpha \bar\psi \gamma^b \partial_{a_\alpha} \lambda 
\prod_\beta \bar\psi \gamma^c \partial_{a_\beta} \partial_b \lambda 
\prod_\gamma \partial_{a_\gamma} \partial_c \lambda \right. 
\nonu
\A \A 
\left. + {1 \over 2} 
\prod_\alpha \bar\psi \gamma^b \partial_{a_\alpha} \lambda 
\prod_\beta \bar\psi \gamma^c \partial_{a_\beta} \lambda 
\prod_\gamma \partial_{a_\gamma} \partial_b \partial_c \lambda 
\right) 
\nonu
\A \A 
- \sum_{n \ge 1} i^{n+1} \kappa^{{n \over 2} + 6} \gamma^A 
\sum_{k_1+\cdots+k_4=n} {n! \over {k_1! \cdots k_4!}} 
\nonu
\A \A 
\times \left( 
\prod_\alpha \bar\psi \gamma^b \partial_{a_\alpha} \lambda 
\prod_\beta \bar\psi \gamma^c \partial_{a_\beta} \partial_b \lambda 
\prod_\gamma \bar\psi \gamma^d \partial_{a_\gamma} \partial_c \lambda 
\prod_\delta \partial_{a_\delta} \partial_d \lambda \right. 
\nonu
\A \A 
+ {1 \over 2} 
\prod_\alpha \bar\psi \gamma^b \partial_{a_\alpha} \lambda 
\prod_\beta \bar\psi \gamma^c \partial_{a_\beta} \lambda 
\prod_\gamma \bar\psi \gamma^d \partial_{a_\gamma} \partial_b \partial_c \lambda 
\prod_\delta \partial_{a_\delta} \partial_d \lambda 
\nonu
\A \A 
+ \prod_\alpha \bar\psi \gamma^b \partial_{a_\alpha} \lambda 
\prod_\beta \bar\psi \gamma^c \partial_{a_\beta} \lambda 
\prod_\gamma \bar\psi \gamma^d \partial_{a_\gamma} \partial_b \lambda 
\prod_\delta \partial_{a_\delta} \partial_c \partial_d \lambda 
\nonu
\A \A 
\left. 
+ {1 \over 6} 
\prod_\alpha \bar\psi \gamma^b \partial_{a_\alpha} \lambda 
\prod_\beta \bar\psi \gamma^c \partial_{a_\beta} \lambda 
\prod_\gamma \bar\psi \gamma^d \partial_{a_\gamma} \lambda 
\prod_\delta \partial_{a_\delta} \partial_b \partial_c \partial_d \lambda 
\right) 
\nonu
\A \A 
+ \sum_{n \ge 1} i^n \kappa^{{n \over 2} + 8} \gamma^A 
\sum_{k_1+\cdots+k_5=n} {n! \over {k_1! \cdots k_5!}} 
\nonu
\A \A 
\times \left\{ 
\prod_\alpha \bar\psi \gamma^b \partial_{a_\alpha} \lambda 
\prod_\beta \bar\psi \gamma^c \partial_{a_\beta} \partial_b \lambda 
\prod_\gamma \bar\psi \gamma^d \partial_{a_\gamma} \partial_c \lambda 
\prod_\delta \bar\psi \gamma^e \partial_{a_\delta} \partial_d \lambda 
\prod_\epsilon \partial_{a_\epsilon} \partial_e \lambda \right. 
\nonu
\A \A 
+ \prod_\alpha \bar\psi \gamma^b \partial_{a_\alpha} \lambda 
\prod_\beta \bar\psi \gamma^c \partial_{a_\beta} \lambda 
\prod_\gamma \bar\psi \gamma^d \partial_{a_\gamma} \partial_b \lambda 
\prod_\delta \bar\psi \gamma^e \partial_{a_\delta} \partial_c \partial_d \lambda 
\prod_\epsilon \partial_{a_\epsilon} \partial_e \lambda 
\nonu
\A \A 
+ \prod_\alpha \bar\psi \gamma^b \partial_{a_\alpha} \lambda 
\prod_\beta \bar\psi \gamma^c \partial_{a_\beta} \lambda 
\prod_\gamma \bar\psi \gamma^d \partial_{a_\gamma} \partial_b \lambda 
\prod_\delta \bar\psi \gamma^e \partial_{a_\delta} \partial_d \lambda 
\prod_\epsilon \partial_{a_\epsilon} \partial_c \partial_e \lambda 
\nonu
\A \A 
+ {1 \over 2} \left( 
\prod_\alpha \bar\psi \gamma^b \partial_{a_\alpha} \lambda 
\prod_\beta \bar\psi \gamma^c \partial_{a_\beta} \lambda 
\prod_\gamma \bar\psi \gamma^d \partial_{a_\gamma} \partial_b \partial_c \lambda 
\prod_\delta \bar\psi \gamma^e \partial_{a_\delta} \partial_d \lambda 
\prod_\epsilon \partial_{a_\epsilon} \partial_e \lambda \right. 
\nonu
\A \A 
+ \prod_\alpha \bar\psi \gamma^b \partial_{a_\alpha} \lambda 
\prod_\beta \bar\psi \gamma^c \partial_{a_\beta} \lambda 
\prod_\gamma \bar\psi \gamma^d \partial_{a_\gamma} \partial_b \lambda 
\prod_\delta \bar\psi \gamma^e \partial_{a_\delta} \partial_c \lambda 
\prod_\epsilon \partial_{a_\epsilon} \partial_d \partial_e \lambda 
\nonu
\A \A 
+ \prod_\alpha \bar\psi \gamma^b \partial_{a_\alpha} \lambda 
\prod_\beta \bar\psi \gamma^c \partial_{a_\beta} \lambda 
\prod_\gamma \bar\psi \gamma^d \partial_{a_\gamma} \lambda 
\prod_\delta \bar\psi \gamma^e \partial_{a_\delta} \partial_b \partial_c \lambda 
\prod_\epsilon \partial_{a_\epsilon} \partial_d \partial_e \lambda 
\nonu
\A \A 
+ \left. \prod_\alpha \bar\psi \gamma^b \partial_{a_\alpha} \lambda 
\prod_\beta \bar\psi \gamma^c \partial_{a_\beta} \lambda 
\prod_\gamma \bar\psi \gamma^d \partial_{a_\gamma} \lambda 
\prod_\delta \bar\psi \gamma^e \partial_{a_\delta} \partial_b \lambda 
\prod_\epsilon \partial_{a_\epsilon} \partial_c \partial_d \partial_e \lambda 
\right) 
\nonu
\A \A 
+ {1 \over 6} 
\prod_\alpha \bar\psi \gamma^b \partial_{a_\alpha} \lambda 
\prod_\beta \bar\psi \gamma^c \partial_{a_\beta} \lambda 
\prod_\gamma \bar\psi \gamma^d \partial_{a_\gamma} \lambda 
\prod_\delta \bar\psi \gamma^e \partial_{a_\delta} \partial_b \partial_c \partial_d \lambda 
\prod_\epsilon \partial_{a_\epsilon} \partial_e \lambda 
\nonu
\A \A 
+ \left. {1 \over 24} 
\prod_\alpha \bar\psi \gamma^b \partial_{a_\alpha} \lambda 
\prod_\beta \bar\psi \gamma^c \partial_{a_\beta} \lambda 
\prod_\gamma \bar\psi \gamma^d \partial_{a_\gamma} \lambda 
\prod_\delta \bar\psi \gamma^e \partial_{a_\delta} \lambda 
\prod_\epsilon \partial_{a_\epsilon} \partial_b \partial_c \partial_d \partial_e \lambda 
\right\}. 
\label{tilde-e1}
\ea
By straightforward calculations we can prove that in all orders 
the fields $\tilde \lambda_0(\psi)$ and $\tilde \lambda_1(\psi)$ 
transform homogeneously under the NL SUSY transformations 
(\ref{NLSUSY}) and (\ref{NLSUSY2}), i.e., 
\ba
\A \A 
\delta_\zeta \tilde \lambda_0(\psi) 
= \xi^a \partial_a \tilde \lambda_0(\psi), 
\nonu
\A \A 
\delta_\zeta \tilde \lambda_1(\psi) 
= \xi^a \partial_a \tilde \lambda_1(\psi), 
\ea
so that $\delta_\zeta \tilde \lambda(\psi) = \xi^a \partial_a \tilde \lambda(\psi)$. 
Therefore, the constraints, 
\be
\tilde \lambda(\psi) 
= \tilde \lambda_0(\psi) + \tilde \lambda_1(\psi) = 0, 
\label{const}
\ee
are NL SUSY invariant and those explicitly give NL SUSY invariant relations 
which connect our NL SUSY actions (\ref{NLSUSYact}) including apparently 
(pathological) higher derivative terms of $\lambda$ 
with the V-A action (\ref{VAact}) described by $\psi$. 
\footnote{
Note that if we consider the NL SUSY invariant constraint, 
$\tilde \lambda_0(\psi) = 0 $, 
then it gives the unique solution $\psi = \lambda$ 
by means of Eq.(\ref{tilde-e0}). 
This is just the case, $[\ {\rm higher\ derivative\ terms\ of\ } \lambda\ ] = 0$ 
in Eq.(\ref{NLSUSYact}), 
i.e., $S(\lambda) = S_{\rm VA}(\lambda) = S_{\rm VA}(\psi)$.}

Here we briefly discuss on a simple example of the NL SUSY invariant 
relations derived from the constraints (\ref{const}). 
Indeed, for the case of $n = 1$ in the above discussion, 
i.e., for the case where the terms up to the first-order derivative of $\lambda$ 
are considered in the fields (\ref{fields}), 
the composite fields $\tilde \lambda_1(\psi)$ of Eq.(\ref{tilde-1}) are 
\be
\tilde \lambda_1(\psi) 
= \left( 1 + \delta_\zeta + {1 \over 2!} \delta_\zeta^2 
+ {1 \over 3!} \delta_\zeta^3 + {1 \over 4!} \delta_\zeta^4 \right) 
\ i \kappa^{1 \over 2} \!\!\not\!\partial \lambda 
\ \ \vert_{\zeta \rightarrow - \kappa \psi}, 
\label{tilde-1ex}
\ee
and the constraints (\ref{const}) have the following form at leading orders, 
\ba
\tilde \lambda(\psi) 
= \A \A \lambda - \psi + i \kappa^{1 \over 2} \ \!\!\not\!\partial \lambda 
+ i \kappa^2 \ \bar\psi \gamma^a \lambda \ \partial_a \lambda 
\nonu
\A \A 
- \kappa^{5 \over 2} 
(\bar\psi \gamma^a \partial_b \lambda \ \gamma^b \partial_a \lambda 
+ \bar\psi \gamma^a \lambda \ \partial_a \!\!\not\!\partial \lambda) 
\nonu
\A \A 
- \kappa^4 \left( \bar\psi \gamma^a \lambda \ \bar\psi \gamma^b \partial_a \lambda 
\ \partial_b \lambda 
+ {1 \over 2} \bar\psi \gamma^a \lambda 
\ \bar\psi \gamma^b \lambda \ \partial_a \partial_b \lambda 
\right) 
\nonu
\A \A 
- i \kappa^{9 \over 2} 
\left( \bar\psi \gamma^a \partial_c \lambda 
\ \bar\psi \gamma^b \partial_a \lambda 
\ \gamma^c \partial_b \lambda 
+ \bar\psi \gamma^a \lambda 
\ \bar\psi \gamma^b \partial_a \partial_c \lambda 
\ \gamma^c \partial_b \lambda \right. 
\nonu
\A \A 
+ \bar\psi \gamma^a \lambda 
\ \bar\psi \gamma^b \partial_a \lambda 
\ \partial_b \!\!\not\!\partial \lambda 
+ \bar\psi \gamma^a \partial_c \lambda 
\ \bar\psi \gamma^b \lambda 
\ \gamma^c \partial_a \partial_b \lambda 
\nonu
\A \A 
\left. + {1 \over 2} \bar\psi \gamma^a \lambda 
\ \bar\psi \gamma^b \lambda 
\ \partial_a \partial_b \!\!\not\!\partial \lambda \right) 
+ {\cal O}(\kappa^6) 
\nonu
\A \A 
= 0. 
\label{fconst1}
\ea
Solving Eq.(\ref{fconst1}) with respect to $\psi$ as a function of $\lambda$ 
gives the NL SUSY invariant relation, 
\ba
\psi 
= \A \A \lambda + i \kappa^{1 \over 2} \!\!\not\!\partial \lambda 
+ \kappa^{5 \over 2} 
(\bar\lambda \gamma^a \!\!\not\!\partial \lambda \ \partial_a \lambda 
- \bar\lambda \gamma^a \partial_b \lambda \ \gamma^b \partial_a \lambda) 
\nonu
\A \A 
+ i \kappa^3 (\ \!\!\not\!\partial \bar\lambda \gamma^a \partial_b \lambda 
\ \gamma^b \partial_a \lambda 
+ \bar\lambda \gamma^a \!\!\not\!\partial \lambda 
\ \partial_a \!\!\not\!\partial \lambda) 
\nonu
\A \A 
+ i \kappa^{9 \over 2} 
(\bar\lambda \gamma^a \!\!\not\!\partial \lambda 
\ \bar\lambda \gamma^b \partial_a \lambda 
\ \partial_b \lambda 
+ \bar\lambda \gamma^a \partial_c \lambda 
\ \bar\lambda \gamma^b \partial_a \lambda 
\ \gamma^c \partial_b \lambda) 
+ {\cal O}(\kappa^5). 
\label{fconst2}
\ea
Substituting this into the V-A action (\ref{VAact}), 
we have the following NL SUSY action, 
\be
S(\lambda) = S_{\rm VA}(\lambda) + \int d^4 x 
\left[ \kappa^{1 \over 2} \partial_a \bar\lambda \partial^a \lambda 
+ {i \over 2} \kappa \partial_a \bar\lambda \gamma^a \Box \lambda \right] 
+ {\cal O}(\lambda^4) 
\label{hdact}
\ee
except for total derivative terms. 
It can be understood from the higher derivative terms 
in ${\cal O}(\lambda^2)$ of the action 
that Eq.(\ref{hdact}) includes apparently a (Weyl) ghost field 
in the higher derivative fermionic field theory (for example, see \cite{Vi}). 
However, those terms in ${\cal O}(\lambda^2)$ do not alter 
the pole structure of the N-G fermion for the on-shell amplitudes 
because Eq.(\ref{hdact}) is equivalent to the standard V-A action 
$S_{\rm VA}(\psi)$ of Eq.(\ref{VAact}) through the NL SUSY invariant 
relation (\ref{fconst2}). 
Eqs.(\ref{fconst2}) and (\ref{hdact}) are 
just the relations as discussed in \cite{ST} by heuristic arguments. 

In the same way, 
for more general cases of $n \ge 2$ 
where the terms up to the second-order and the higher derivatives 
of $\lambda$ are considered in the fields (\ref{fields}) 
and in $\tilde \lambda_1(\psi)$ of Eq.(\ref{tilde-1}), 
the constraints (\ref{const}) can be solved unambiguously 
with respect to $\psi$ as a function of $\lambda$ in all orders. 
By substituting the obtained general NL SUSY invariant relations 
as, at ${\cal O}(\lambda^1)$, 
\ba
\psi \A = \A \lambda 
+ \sum_{n \ge 1} \{ (i \kappa^{1 \over 2})^n 
\ \!\!\not\!\partial^n \lambda + {\cal O}(\kappa^{{n \over 2} + 2}) \} 
\nonu
\A = \A \lambda + i \kappa^{1 \over 2} \ \!\!\not\!\partial \lambda 
- \kappa \ \!\!\not\!\partial \!\!\not\!\partial \lambda + \cdots 
+ \sum_{n \ge 1} {\cal O}(\kappa^{{n \over 2} + 2}) 
\label{g-expand}
\ea
into the V-A action (\ref{VAact}), 
we have also the corresponding NL SUSY actions including apparently 
(pathological) higher derivative terms of $\lambda$, 
\ba
S(\lambda) = \A \A S_{\rm VA}(\lambda) + \int d^4 x 
\left[ {i \over 2} \sum_{n \ge 1} (i \kappa^{1 \over 2})^n 
\{ \bar\lambda \!\!\not\!\partial^{n+1} \lambda 
+ (-)^n \!\!\not\!\partial^n \bar\lambda \!\!\not\!\partial \lambda \} 
\right. 
\nonu
\A \A 
\left. + {i \over 2} \sum_{n \ge 1, m = 1, ..., n } 
(-)^n (i \kappa^{1 \over 2})^{n+m} 
\!\!\not\!\partial^n \bar\lambda \!\!\not\!\partial^{m+1} \lambda 
\right] + {\cal O}(\lambda^4) 
\nonu
= \A \A S_{\rm VA}(\lambda) + \int d^4 x 
\left[ - {1 \over 2} \kappa^{1 \over 2} 
(\bar\lambda \Box \lambda 
- \partial_a \bar\lambda \gamma^a \!\!\not\!\partial \lambda) 
\right. 
\nonu
\A \A 
+ {i \over 2} \kappa 
(\partial_a \bar\lambda \gamma^a \Box \lambda 
- \bar\lambda \Box \!\!\not\!\partial \lambda 
- \Box \bar\lambda \!\!\not\!\partial \lambda) 
\nonu
\A \A 
\left. - {1 \over 2} \kappa^{3 \over 2} 
(\ \!\!\not\!\partial \bar\lambda \Box \!\!\not\!\partial \lambda 
- \Box \bar\lambda \Box \lambda) 
+ {i \over 2} \kappa^2 \Box \bar\lambda \Box \!\!\not\!\partial \lambda 
+ \cdots \ \right] + {\cal O}(\lambda^4) 
\label{g-hdact}
\ea
which are equivalent to the standard V-A action. 
Our results, e.g., Eqs.(\ref{g-expand}) and (\ref{g-hdact}), 
are also valid for the arbitrary coefficients $c_n$ 
in Eq.(\ref{fields}). 

To conclude, we have extended the arguments in \cite{ST} 
with respect to the universality of the NL SUSY actions with the N-G fermion 
to more general cases where $\psi$ in (\ref{VAact}) is expanded 
in terms of $\lambda$ in (\ref{NLSUSYact}) and its higher derivatives 
at ${\cal O}(\lambda^1)$ as Eq.(\ref{g-expand}). 
In order to determine higher order terms of Eq.(\ref{g-expand}) 
in NL SUSY invariant way, 
according to the algorithmic procedure given in \cite{Iv}, 
we have found the composite fields $\tilde \lambda(\psi)$ defined by the sum 
of Eqs.(\ref{tilde-0}) and (\ref{tilde-1}) 
(or explicitly, Eqs.(\ref{tilde-e0}) and (\ref{tilde-e1})) 
which transform homogeneously under the NL SUSY transformations 
(\ref{NLSUSY}) and (\ref{NLSUSY2}). 
Consequently, we have obtained the constraints (\ref{const}) 
which connect our NL SUSY actions (\ref{NLSUSYact}) 
with the V-A action (\ref{VAact}). 
In our previous work in \cite{ST}, 
we have constructed a NL SUSY invariant relation by heuristic arguments 
for the case where $\psi$ is expanded 
up to the first-order derivative of $\lambda$ at ${\cal O}(\lambda^1)$, 
i.e., $\psi = \lambda + i \kappa^{1 \over 2} \ \!\not\!\partial \lambda 
+ {\cal O}(\kappa^{5 \over 2})$. 
We have proved in this letter that the NL SUSY invariant relation 
is derived in Eq.(\ref{fconst2}) as a solution 
of the constraint (\ref{fconst1}). 
We have also discussed in Eqs.(\ref{g-expand}) and (\ref{g-hdact}) 
that more general NL SUSY invariant relations and the NL SUSY actions 
including apparently (pathological) higher derivatives of the N-G fermion 
can be obtained by solving the constraints (\ref{const}) 
with respect to $\psi$ as a function of $\lambda$ 
and by substituting the solutions into the standard V-A action (\ref{VAact}).

%
%

\newpage

%
\newcommand{\NP}[1]{{\it Nucl.\ Phys.\ }{\bf #1}}
\newcommand{\PL}[1]{{\it Phys.\ Lett.\ }{\bf #1}}
\newcommand{\CMP}[1]{{\it Commun.\ Math.\ Phys.\ }{\bf #1}}
\newcommand{\MPL}[1]{{\it Mod.\ Phys.\ Lett.\ }{\bf #1}}
\newcommand{\IJMP}[1]{{\it Int.\ J. Mod.\ Phys.\ }{\bf #1}}
\newcommand{\PR}[1]{{\it Phys.\ Rev.\ }{\bf #1}}
\newcommand{\PRL}[1]{{\it Phys.\ Rev.\ Lett.\ }{\bf #1}}
\newcommand{\PTP}[1]{{\it Prog.\ Theor.\ Phys.\ }{\bf #1}}
\newcommand{\PTPS}[1]{{\it Prog.\ Theor.\ Phys.\ Suppl.\ }{\bf #1}}
\newcommand{\AP}[1]{{\it Ann.\ Phys.\ }{\bf #1}}


\begin{thebibliography}{100}

\bibitem{VA}
D. V. Volkov and V. P. Akulov,  
{\it JETP Lett.} {\bf 16} (1972) 438; 
\PL{B46} (1973) 109. 

\bibitem{SS}
A. Salam and J. Strathdee, \PL{B49} (1974) 465. 

\bibitem{FI}
P. Fayet and J. Iliopoulos, \PL{B51} (1974) 461. 

\bibitem{O}
L. O'Raifeartaigh, \NP{B96} (1975) 331. 

\bibitem{WZ}
J. Wess and B. Zumino, \PL{B49} (1974) 52; \NP{B78} (1974) 1. 

\bibitem{Fa}
P. Fayet, \NP{B113} (1976) 135. 

\bibitem{IK}
E. A. Ivanov and A. A. Kapustnikov, 
Relation between linear and nonlinear realizations of 
supersymmetry, JINR Dubna Report No.\ E2-10765, 1977, unpublished; \\
E. A. Ivanov and A. A. Kapustnikov, {\it J. Phys.\ }{\bf A11} (1978) 2375; 
{\it J. Phys.\ }{\bf G8} (1982) 167. 

\bibitem{Ro}
M. Ro\v{c}ek, \PRL{41} (1978) 451. 

\bibitem{UZ}
T. Uematsu and C.K. Zachos, \NP{B201} (1982) 250. 

\bibitem{STT}
K. Shima, Y. Tanii and M. Tsuda, \PL{B525} (2002) 183; 
\PL{B546} (2002) 162. 

\bibitem{KSST}
K. Shima, {\it Z. Phys.} {\bf C18} (1983) 25; 
{\it European Phys. J.} {\bf C7} (1999) 341; \\
K. Shima, \PL{B501} (2001) 237; \\
K. Shima and M. Tsuda, \PL{B507} (2001) 260; 
{\it Class. Quantum Grav.} {\bf 19} (2002) 5101. 

\bibitem{SW}
S. Samuel and J. Wess, \NP{B221} (1983) 153. 

\bibitem{HK}
T. Hatanaka and S. V. Ketov, \PL{B580} (2004) 265. 

\bibitem{ST}
K. Shima and M. Tsuda, 
{\it Mod. Phys. Lett.} {\bf A19} (2004) 1357. 

\bibitem{Iv}
E. A. Ivanov and A.A. Kapustnikov, {\it Theor. Math. Phys.\ }{\bf 129} (2001) 1543. 

\bibitem{Vi}
E. J. S. Villase\~{n}or, 
{\it J. Phys.} {\bf A35} (2002) 6169. 

\end{thebibliography}
\end{document}